\documentstyle[preprint,eqsecnum,aps]{revtex}
\tighten
\newcommand{\beq}{\begin{equation}}
\newcommand{\beqa}{\begin{eqnarray}}
\newcommand{\eeq}{\end{equation}}
\newcommand{\eeqa}{\end{eqnarray}}

\newcommand{\siml}{\lesssim}

\begin{document}
\draft

\preprint{YITP-97-3, gr-qc/9707043}

\title{
Anisotropy of the Cosmic Background Radiation
implies the Violation of
the Strong Energy Condition in Bianchi type I Universe
}

\author{
Takeshi Chiba, Shinji Mukohyama, and Takashi Nakamura
}
\address{Yukawa Institute for Theoretical Physics, Kyoto University, 
Kyoto 606-01, Japan}

\date{\today}

\maketitle

\vspace{1cm}

\begin{abstract}
We consider the horizon problem in a homogeneous but anisotropic 
universe (Bianchi type I). 
We show that the problem cannot be solved if 
(1) the matter obeys the strong energy condition 
with the positive energy density
and (2) the Einstein equations hold. 
The strong energy condition is violated during cosmological inflation.
\end{abstract}


\section{introduction}

The discovery of the cosmic microwave background (CMB)\cite{pw} verified 
the hot big bang cosmology. The high degree of its 
isotropy\cite{cobe}, however, 
gave rise to the horizon problem: Why could causally disconnected regions be 
isotropized? The inflationary universe scenario\cite{inf} may solve 
the problem because inflation made it possible for the universe to expand
enormously up to the presently observable scale in a very short time. 
However inflation is the sufficient condition even if the cosmic no
hair conjecture\cite{cnc} is proved.  Here, a problem again arises:
 Is inflation
the unique solution to the horizon problem? What is the general
requirement for the solution of the horizon problem?

Recently, Liddle showed that in FRW universe the horizon problem
cannot be solved without violating the strong energy
condition if gravity can be treated classically\cite{liddle}. 
Actually the strong energy condition is violated during inflation. 
The generalization of  his result to a more general inhomogeneous and 
anisotropic universe is urgent. The motivations 
are two folds. (1) The universe around the Planck epoch is expected to be
highly inhomogeneous and anisotropic.  
(2) Even from the Planck epoch afterwards, the universe may be highly 
inhomogeneous, because it may experience many stages of phase transition,
such as GUT, electroweak, quark-hadron, etc. 
Since the particle horizon from the Planck time to 
the time of nucleosynthesis is
essential to Liddle's argument, we need to generalize his argument to
an inhomogeneous  and anisotropic universe. 
In short we are
concerned about the ``structural stability'' of Liddle's argument. 
That is, is his result specific to FRW universe, or does it hold quite 
generally?

Maartens, Ellis, and Stoeger\cite{ellis} 
recently showed that if the residual dipole of the cosmic microwave
background anisotropy vanishes to first order of perturbations and if
the quadrupole and the octapole are spatially homogeneous to first
order, then the spacetime is locally Bianchi I to first order since
the decoupling to the present day. Therefore as a first step towards 
the general case, we shall consider  
a homogeneous but anisotropic universe (Bianchi type I) in this Letter. 

In Sec.2, we compute the comoving Hubble distance in a homogeneous but 
anisotropic universe. We find that the horizon problem again cannot be
solved without violating the strong energy condition. 
In Sec.3, we give an argument generalizing to an inhomogeneous
universe. 
Sec.4 is devoted to summary.

\section{horizon problem in homogeneous but anisotropic universe}
		\label{sec:Bianchi}
The Bianchi I universe is described by the following metric
\beq
ds^2=-dt^2 +X^2(t)dx^2+Y^2(t)dy^2+Z^2(t)dz^2.
\eeq
We normalize each scale factor such that $X(t)=Y(t)=Z(t)=1$ at the present
time ($t_0$). We define the averaged scale factor by $R^3(t)=X(t)Y(t)Z(t)$. 
We assume that the energy-momentum tensor of the matter obeys
\beq
T_{\mu \nu}=(\rho + p)u_\mu u_\nu +pg_{\mu \nu}.
\eeq
 From the Einstein equations we have
\beqa
{\dot X\over X}&=&{\dot R\over R}+{\sigma_x\over R^3},\label{X-eq}\\
{\dot Y\over Y}&=&{\dot R\over R}+{\sigma_y\over R^3},\label{Y-eq}\\
{\dot Z\over Z}&=&{\dot R\over R}+{\sigma_z\over R^3},\label{Z-eq}\\
\sigma_x+\sigma_y&+&\sigma_z=0,	\label{sigma}\\
H^2\equiv\left({\dot R\over R}\right)^2&=&{8\pi G \over 3}\rho+
{\sigma_0^2\over 6R^6},\label{hubble} \\
\sigma_0^2&=& \sigma_x^2+\sigma_y^2+\sigma_z^2,	\label{sigma^2}
\eeqa
where  dots denote the time derivative and  $\sigma_i$ is
a constant representing the present shear in $i$th-direction.
The Bianchi identify reads
\beq
(\rho R^3)^{.}+p(R^3)^{.}=0.
\eeq
The anisotropy of cosmic microwave background seen by COBE sets 
a limit on $\sigma_0$\cite{ms}
\beq
{\sigma_0\over 3H_0} \leq 6.9\times 10^{-10}.
\label{cobe}
\eeq 
Without loss of generality we can impose the condition 
\begin{equation}
 \sigma_x \geq \sigma_y,\ \sigma_x \geq \sigma_z. \label{sigma-x}
\end{equation}
With the help of Eq. (\ref{X-eq})-(\ref{Z-eq}) these conditions imply 
\begin{equation}
 X(t)\leq Y(t),\ X(t)\leq Z(t) \quad (t\leq t_0 ),
\end{equation}
since we have normalized each scale factor such that 
$X(t_0)=Y(t_0)=Z(t_0)=1$. 
Thus, for $t\leq t_0$, 
\begin{equation}
 (dx^2+dy^2+dz^2)^{1/2}\leq\frac{dt}{X(t)}
\end{equation}
along any null lines. This means that the comoving distance
 $d_{comm}$ along an
arbitrary null line from $t=t_1$ to $t=t_2$ ($t_{pl}\leq t_1<t_2\leq
t_0$), which can be regarded as the communication distance in
the terminology of Liddle, is bounded above as
\begin{equation}
 d_{comm}(t_1,t_2)\equiv \int _{{\rm along~ null}} \sqrt{dx^2+dy^2+dz^2} 
\leq\int^{t_2}_{t_1}\frac{dt}{X(t)}.
\end{equation}

Now we have only to consider $X(t)$ among $X(t)$,
$Y(t)$ and $Z(t)$ to show that $d_{comm}(t_{pl},t) \ll 1/H_0$ below. 
 First let us
define an effective density $\rho_x$ and an effective pressure $p_x$
by 
\begin{eqnarray}
 \left(\frac{\dot{X}}{X}\right)^2 & = & \frac{8\pi G}{3}\rho_x,\label{hx1}	\\
 \dot{\rho_x} & = & -3(\rho_x+p_x)\frac{\dot{X}}{X}.
\label{hx2}
\end{eqnarray}
Using $x-x$ component of the evolution equation of the extrinsic
curvatures, we have
\begin{eqnarray}
 \rho_x & = & \rho + \frac{\sigma_0^2}{16\pi GR^6} 
	+ \frac{3\sigma_x^2}{8\pi GR^6} 
	+ \frac{3\sigma_x\dot{R}}{4\pi GR^4},\label{rhox}\\
 p_x & = & p + \frac{\sigma_0^2}{16\pi GR^6}
	- \frac{3\sigma_x^2}{8\pi GR^6},
\end{eqnarray}
and 
\begin{equation}
 \rho_x+3p_x = (\rho +3p)+\frac{1}{4\pi GR^6}\left(
	\sigma_0^2-3\sigma_x^2+3\sigma_xR^2\dot{R}\right).
\end{equation}

Now let us assume
$\rho\geq 0$.  From Eq. (\ref{sigma-x}), (\ref{sigma}) and
(\ref{sigma^2}), $\sigma_x$ is
bounded from below and above as 
$0\leq\sigma_x\leq\sqrt{6}|\sigma_0|/3$. Next using the positivity
of the density , 
$R^2\dot{R}\geq |\sigma_0|/\sqrt{6}$ is derived from
Eq. (\ref{hubble}). Then we can prove that
$\rho_x+3p_x$ is greater than $\rho +3p$ as 
\begin{eqnarray}
 (\rho_x+3p_x)-(\rho +3p) & \geq & 
	\frac{1}{4\pi GR^6}\left(\sigma_0^2-3\sigma_x^2+
	\frac{\sqrt{6}}{2}\sigma_x|\sigma_0|\right)	\\
	& = & \frac{1}{4\pi GR^6}
	\left(|\sigma_0|-\frac{\sqrt{6}}{2}\sigma_x\right)
	\left(|\sigma_0|+\sqrt{6}\sigma_x\right)	\\
	& \geq & 0.
\end{eqnarray}
We thus finally have the strong energy condition for $\rho_x$ 
and $p_x$ as
\begin{equation}
 \rho_x+3p_x \geq \rho+3p \geq 0. 
\end{equation}
provided that the original version of the strong energy condition holds 
($\rho +3p\geq 0$).

 Now under the above strong energy condition  we shall prove 
the relation as 
\beq
d_{comm}(t_{pl},t)\leq 
\int_{t_{pl}}^{t}{dt'\over X(t')} << {1\over H_0},
\label{dcomm}
\eeq
where $t$ may be taken to be 
 any time between the Planck time and the decoupling time. 
We take the standpoint that the matter content of the Universe 
is well understood after the big bang nucleosynthesis. 
  In order to 
 evaluate the integral in Eq.(\ref{dcomm}),
 we divide the time range into  two  epochs: 
(1) from the Planck time $t_{pl}$ to 
 the time of nucleosynthesis $t_{nuc}$ which is
defined by the time when the neutron-to-proton ratio is frozen out; 
(2) from $t_{nuc}$ to the time of the decoupling of the microwave
background $t_{dec}$. In epoch (1), we assume that
 matter obeys the energy condition such that 
\beq
 \rho_x+3p_x \geq 0, 
\eeq
which has been derived under the strong energy condition
with positive energy density. 
In epoch (2), the universe is dominated by the ordinary dust matter
and radiation.

\subsection{from $t_{pl}$ to $t_{nuc}$}
From Eqs.(\ref{hx1}) and (\ref{hx2}) we have
\beq
\frac{d \ln X}{d \ln \rho_x}=-\frac{1}{3}\frac{\rho_x}{\rho_x+p_x},
\eeq
which clearly shows that $X$ decrease most rapidly when the pressure
$p_x$ is the lowest $p_x=-\rho_x/3$. 
The integral of Eq.(\ref{dcomm}) can be rewritten as
\beqa
\int_{t_{pl}}^{t_{nuc}}{dt'\over
X(t')}&=&-\int_{(\rho_x)_{pl}}^{(\rho_x)_{nuc}}
{d\rho_x\over 3H_xX(\rho_x)(\rho_x+p_x)}\nonumber\\
&=&-{1\over
\sqrt{24\pi G}}\int_{(\rho_x)_{pl}}^{(\rho_x)_{nuc}}{d\rho_x\over
X(\rho_x+p_x)\sqrt{\rho_x}},
\eeqa
where
\beq
H_x\equiv \frac{\dot{X}}{X}
\eeq
The integral is maximized by the lowest pressure\cite{liddle} and we have
\beq
d_{comm}(t_{pl},t_{nuc})\leq {1\over 2X_{nuc}(H_x)_{nuc}}
\ln{(\rho_x)_{pl}\over (\rho_x)_{nuc}}.
\label{planck}
\eeq
From the definition of $\rho_x$ in Eq.(\ref{rhox}),
$(\rho_x)_{pl}$ is maximized when $\sigma_0, \sigma_x$ and
$\dot{R}$ are maximized and 
$R_{pl}$ is minimized. $R_{pl}$ is minimized by the lowest
possible pressure $p_x=-\rho_x/3$. Since $t_{pl} \sim 10^{-43}{\rm sec}$ and
$t_{nuc} \sim 1{\rm sec}$, we have 
\beqa
 (\rho_x)_{pl} & = & \rho_{pl} + \frac{\sigma_0^2}{16\pi GR^6} 
	+ \frac{3\sigma_x^2}{8\pi GR^6} 
	+ \frac{3\sigma_x\dot{R}}{4\pi GR^4}	
                                  \nonumber\\
& \siml & \rho_{pl}+{9\over 16\pi G}{\sigma_0^2\over
( R_{pl})^6}\nonumber\\
& \siml & \rho_{pl}+{9\over 16\pi G}{(10^{-9}\times H_0)^2\over
(10^{-32})^6},
\eeqa
where $\rho_{pl} \sim (10^{19}{\rm Gev})^4$ is the Planck energy. 
Thus we can estimate the right hand side of Eq.(\ref{planck}) as
\beq
\ln{(\rho_x)_{pl}\over (\rho_x)_{nuc}}   <  620,
\eeq
where $(\rho_x)_{nul} \sim (10^{-3}{\rm Gev})^4$.  
Since $H^{-1}_{nuc}/R_{nuc}\simeq 10^{-4}{\rm Mpc}$, 
we find that at most
\beq
 d_{comm}(t_{pl},t_{nuc}) < 1\times 10^{-5}/H_0.
\eeq

\subsection{from $t_{nuc}$ to $t_{dec}$}

Since the shear is negligible in this epoch, 
the analysis is completely the same as  Liddle's\cite{liddle}. Here we 
repeat his analysis for completeness. 
Between nucleosynthesis and decoupling, the universe is either in the
stage of radiation-dominant or matter-dominant and the distance is
bounded above by assuming matter is dominated throughout
\beq
d_{comm}(t_{nuc},t_{dec}) \leq {2\over R_{dec}H_{dec}}.
\eeq
Since $H^{-1}_{dec}/R_{dec}\simeq 100{\rm Mpc}$, we find  
\beq
d_{comm}(t_{nuc},t_{dec}) \leq 6.6 \times 10^{-2}/H_0.
\eeq

To conclude, we cannot have 
$d_{comm}(t_{pl},t_{dec})  > 1/H_0$ in a homogeneous but anisotropic
universe. 

\section{ generalizing to inhomogeneous universes}

In the previous section  
we have shown that  the horizon problem is not solved under the
strong energy condition even  in a homogeneous but 
anisotropic universe considering the current 
limit on the anisotropy of the universe set by CMB.
Therefore the strong energy condition should have been violated in the
early universe. However a natural question arises: 
How can we generalize the above result to an inhomogeneous universe?
Here we show that if the concept of the averaged scale factor makes
sense, we can prove the ``no-go theorem''. 
More precisely we show the following:

{\bf Theorem}

{\it If we assume that 
\begin{itemize}
\item{1)} the universe can be foliated by geodesic slicing, 
\item{2)} matter satisfies the strong energy condition 
$(T_{ab}-{1\over 2}g_{ab}T)n^an^b \geq 0$,  
\item{3)} the spatial curvature is everywhere not positive  
$~^3R\leq 0$,
\item{4)} $~^3R$ and $\sigma_{ab}\sigma^{ab}$ take the value of order 
the Planck scale at $t_{pl}$,
\end{itemize} 
then the inequality $\int dt/R(t) < 1/H_0$ follows for 
the averaged scale factor $R$ defined below.} 

We take the geodesic slice, then the following two equations are
necessary in our argument:
\beqa
{2\over 3}K^2&=&-~^3R+\sigma_{ab}\sigma^{ab}+16\pi GT_{ab}n^an^b, \nonumber\\
\dot{K}&=& -{1\over 3}K^2-\sigma_{ab}\sigma^{ab}-8\pi G(T_{ab}-{1\over
2}g_{ab}T)n^an^b, 
\eeqa
where $n^a$ is the unit normal to the spacelike hypersurface, 
$K$ is the trace of the extrinsic curvature $K_{ab}$, and 
the shear tensor $\sigma_{ab}$ is defined by 
\beq
K_{ab}={1\over 2}\dot{h}_{ab}={1\over 3}Kh_{ab}+\sigma_{ab}.
\eeq
$h_{ab}$ is the spatial metric $h_{ab}=g_{ab}+n_an_b$. 
We define the effective Hubble parameter (``volume expansion
rate''\cite{nncs}) $H$ 
and the effective scale factor $R$ 
by 
\beqa
H&=&{1\over 3}K\nonumber\\
{\dot{R}\over R}&=& H.
\eeqa
Similarly, we can define an effective density $\widetilde{\rho}$ and 
an effective pressure $\widetilde{p}$ by
\beqa
H^2&=&{8\pi G\over 3}\widetilde{\rho},\nonumber\\
\dot{\widetilde{\rho}}&=& -3H(\widetilde{\rho}+\widetilde{p}).
\eeqa
$\widetilde{\rho}$ and $\widetilde{p}$ are written as 
\beqa
\widetilde{\rho} &=&T_{ab}n^an^b+{1\over 16\pi G}
(\sigma_{ab}\sigma^{ab} -~^3R),\\
\widetilde{p} &=&{1\over 3}T_{ab}h^{ab}+{1\over 48\pi G}
(~^3R + 3\sigma_{ab}\sigma^{ab}).
\eeqa
We first show that $\widetilde{\rho}$ is positive as
\beq
\widetilde{\rho}=T_{ab}n^an^b+{1\over 16\pi G}(\sigma_{ab}\sigma^{ab}
-~^3R)\geq 
\rho >0.
\eeq
We can also show that the strong energy condition for 
$\widetilde{\rho}$ and $\widetilde{p}$ as
\beq
\widetilde{\rho}+3\widetilde{p}=\rho+T_{ab}h^{ab}+{1\over 4\pi G}
\sigma_{ab}\sigma^{ab}\geq 0,
\eeq
if the strong energy condition ($(T_{ab}-{1\over 2}g_{ab}T)n^an^b 
\geq 0$) is satisfied.
The proof of the theorem may follow by replacing $X, \rho_x$ and $p_x$
in the previous section with  $R, \widetilde{\rho}$ and $\widetilde{p}$,
respectively.

\section{summary}

We have shown that in a  homogeneous but anisotropic 
(Bianchi I) universe the horizon problem  cannot be solved 
if (1) matter satisfies the strong energy condition and (2) the
Einstein equations hold. 
Changing the gravity theory would not change the
result as shown by Liddle\cite{liddle}. 
It  would be very interesting to note that 
the anisotropy of CMB alone may imply that anomalous
phenomena must have happened in the very early stage of the universe: the
violation of the strong energy condition. 

It should be noted that the interpretation of the origin of 
fluctuations of cosmic microwave background is not yet conclusive. 
Topological defects models can generate large angle cosmic microwave
background fluctuations. Such models, however, do not generate
perturbations well above the Hubble radius. Future observations will
distinguish inflation  form  defect models\cite{hsw}. 

It is also noted 
that our analysis as well as Liddle's is within the realm of classical 
theory. We have to keep in mind that the inflation may 
not be the only solution to the horizon problem. 
In fact, the correlation beyond the horizon does exist in any quantum field
theory\cite{wald}. The existence of correlations beyond the horizon 
might have played an important role in the early universe. 
What we have shown here is that in a  homogeneous but anisotropic
universe the horizon problem may be solved either by the causal 
processes during a period of 
inflation or the acausal processes of quantum gravity. 

\acknowledgments
This work was supported in part by a Grant-in-Aid for Basic Research
of Ministry of Education, Culture, Science and Sports (08NP0801).

\end{document}